\documentstyle[12pt]{article}
\newcommand{\nc}{\newcommand}
\nc{\renc}{\renewcommand}

\nc{\etal}{\mbox{\it et al. }}
\nc{\ie}{{\it i.e.}}
\nc{\eg}{{\it e.g.}}

\renc{\thefootnote}{\arabic{footnote}}
\nc{\capt}[1]{{\bf Figure.} {\small\sl #1}}

\nc{\eqs}[2]{\mbox{Eqs.~(\ref{#1},\,\ref{#2})}}
\nc{\eq}[1]{\mbox{Eq.~(\ref{#1})}}

\nc{\figs}[2]{\mbox{Figs.~(\ref{#1},\,\ref{#2})}}
\nc{\fig}[1]{\mbox{Fig~.(\ref{#1})}}

\nc{\tag}[1]{\label{#1} \marginpar{{\footnotesize #1}}}
\nc{\mtag}[1]{\label{#1} \mbox{\marginpar{{\footnotesize #1}}}}
\renc{\baselinestretch}{1.2}
\newlength{\overeqskip}
\newlength{\undereqskip}
\nc{\be}[1]{\begin{equation} \mbox{$\label{#1}$}}
\nc{\bea}[1]{\begin{eqnarray} \mbox{$\label{#1}$}}
\nc{\Section}[2]{\section{#2}\label{#1}}
\nc{\Bibitem}[1]{\bibitem{#1}}
\nc{\Label}[1]{\label{#1}}

\nc{\eea}{\vspace{\undereqskip}\end{eqnarray}}
\nc{\ee}{\vspace{\undereqskip}\end{equation}}
\nc{\bdm}{\begin{displaymath}}
\nc{\edm}{\end{displaymath}}
\nc{\dpsty}{\displaystyle}
\nc{\bc}{\begin{center}}
\nc{\ec}{\end{center}}
\nc{\ba}{\begin{array}}
\nc{\ea}{\end{array}}
\nc{\bab}{\begin{abstract}}
\nc{\eab}{\end{abstract}}
\nc{\btab}{\begin{tabular}}
\nc{\etab}{\end{tabular}}
\nc{\bit}{\begin{itemize}}
\nc{\eit}{\end{itemize}}
\nc{\ben}{\begin{enumerate}}
\nc{\een}{\end{enumerate}}
\nc{\bfig}{\begin{figure}}
\nc{\efig}{\end{figure}}
\nc{\arreq}{&\!=\!&}
\nc{\arrmi}{&\!-\!&}
\nc{\arrpl}{&\!+\!&}
\nc{\arrap}{&\!\!\!\approx\!\!\!&}
\nc{\non}{\nonumber\\*}
\nc{\align}{\!\!\!\!\!\!\!\!&&}

\def\lsim{\; \raise0.3ex\hbox{$<$\kern-0.75em
      \raise-1.1ex\hbox{$\sim$}}\; }
\def\gsim{\; \raise0.3ex\hbox{$>$\kern-0.75em
      \raise-1.1ex\hbox{$\sim$}}\; }
\nc{\DOT}{\hspace{-0.08in}{\bf .}\hspace{0.1in}}
\nc{\Laada}{\hbox {$\sqcap$ \kern -1em $\sqcup$}}
\nc\loota{{\scriptstyle\sqcap\kern-0.55em\hbox{$\scriptstyle\sqcup$}}}
\nc\Loota{{\sqcap\kern-0.65em\hbox{$\sqcup$}}}
\nc\laada{\Loota}
\nc{\qed}{\hskip 3em \hbox{\BOX} \vskip 2ex}

\nc{\real}{{\rm I \! R}}
\nc{\Z}{{\sf Z \!\!\! Z}}
\nc{\complex}{{\rm C\!\!\! {\sf I}\,\,}}
\def\bigid{\leavevmode\hbox{\small1\kern-3.8pt\normalsize1}}
\def\id{\leavevmode\hbox{\small1\kern-3.3pt\normalsize1}}
\nc{\slask}{\!\!\!/}
\nc{\bis}{{\prime\prime}}
\nc{\pa}{\partial}
\nc{\na}{\nabla}
\nc{\ra}{\rangle}
\nc{\la}{\langle}
\nc{\goto}{\rightarrow}
\nc{\swap}{\leftrightarrow}

\nc{\EE}[1]{ \mbox{$\cdot10^{#1}$} }
\nc{\abs}[1]{\left|#1\right|}
\nc{\at}[2]{\left.#1\right|_{#2}}
\nc{\norm}[1]{\|#1\|}
\nc{\abscut}[2]{\Abs{#1}_{\scriptscriptstyle#2}}
\nc{\vek}[1]{{\rm\bf #1}}
\nc{\integral}[2]{\int\limits_{#1}^{#2}}
\nc{\inv}[1]{\frac{1}{#1}}
\nc{\dd}[2]{{{\partial #1}\over{\partial #2}}}
\nc{\ddd}[2]{{{{\partial}^2 #1}\over{\partial {#2}^2}}}
\nc{\dddd}[3]{{{{\partial}^2 #1}\over
	{\partial #2 \partial #3}}}
\nc{\dder}[2]{{{d #1}\over{d #2}}}
\nc{\ddder}[2]{{{d^2 #1}\over{d {#2}^2}}}
\nc{\dddder}[3]{{d^2 #1}\over
	{d #2 d #3}}
\nc{\dx}[1]{d\,^{#1}x}
\nc{\dy}[1]{d\,^{#1}y}
\nc{\dz}[1]{d\,^{#1}z}
\nc{\dl}[1]{\frac{d\,^{#1}l}{(2\pi)^{#1}}}
\nc{\dk}[1]{\frac{d\,^{#1}k}{(2\pi)^{#1}}}
\nc{\dq}[1]{\frac{d\,^{#1}q}{(2\pi)^{#1}}}

\nc{\cc}{\mbox{$c.c.$ }}
\nc{\hc}{\mbox{$h.c.$ }}
\nc{\cf}{cf.\ }
\nc{\erfc}{{\rm erfc}}
\nc{\Tr}{{\rm Tr\,}}
\nc{\tr}{{\rm tr\,}}
\nc{\pol}{{\rm pol}}
\nc{\sign}{{\rm sign}}
\nc{\bfT}{{\bf T }}
\def\GeV{{\rm\ GeV}}

\def\TeV{{\rm\ TeV}}

\nc{\cA}{{\cal A}}
\nc{\cB}{{\cal B}}
\nc{\cD}{{\cal D}}
\nc{\cE}{{\cal E}}
\nc{\cG}{{\cal G}}
\nc{\cH}{{\cal H}}
\nc{\cL}{{\cal L}}
\nc{\cO}{{\cal O}}
\nc{\cT}{{\cal T}}
\nc{\cN}{{\cal N}}
\nc{\rvac}[1]{|{\cal O}#1\rangle}
\nc{\lvac}[1]{\langle{\cal O}#1|}
\nc{\rvacb}[1]{|{\cal O}_\beta #1\rangle}
\nc{\lvacb}[1]{\langle{\cal O}_\beta #1 |}
\nc{\bb}{\bar{\beta}}
\nc{\bt}{\tilde{\beta}}
\nc{\ctH}{\tilde{\cal H}}
\nc{\chH}{\hat{\cal H}}
\nc{\1}{\aa}
\nc{\2}{\"{a}}
\nc{\3}{\"{o}}
\nc{\4}{\AA}
\nc{\5}{\"{A}}
\nc{\6}{\"{O}}
\nc{\al}{\alpha}
\nc{\g}{\gamma}
\nc{\Del}{\Delta}
\nc{\e}{\epsilon}
\nc{\eps}{\epsilon}
\nc{\lam}{\lambda}
\nc{\om}{\omega}
\nc{\Om}{\Omega}
\nc{\ve}{\varepsilon}
\nc{\mn}{{\mu\nu}}
\nc{\k}{\kappa}
\nc{\vp}{\varphi}

\nc{\advp}[3]{{\it  Adv.\ in\ Phys.\ }{{\bf #1} {(#2)} {#3}}}
\nc{\annp}[3]{{\it  Ann.\ Phys.\ (N.Y.)\ }{{\bf #1} {(#2)} {#3}}}
\nc{\apl}[3]{{\it  Appl. Phys. Lett. }{{\bf #1} {(#2)} {#3}}}
\nc{\apj}[3]{{\it  Ap.\ J.\ }{{\bf #1} {(#2)} {#3}}}
\nc{\apjl}[3]{{\it  Ap.\ J.\ Lett.\ }{{\bf #1} {(#2)} {#3}}}
\nc{\app}[3]{{\it Astropart.\ Phys.\ }{{\bf #1} {(#2)} {#3}}}
\nc{\cmp}[3]{{\it  Comm.\ Math.\ Phys.\ }{{ \bf #1} {(#2)} {#3}}}
\nc{\cqg}[3]{{\it  Class.\ Quant.\ Grav.\ }{{\bf #1} {(#2)} {#3}}}
\nc{\epl}[3]{{\it  Europhys.\ Lett.\ }{{\bf #1} {(#2)} {#3}}}
\nc{\ijmp}[3]{{\it Int.\ J.\ Mod.\ Phys.\ }{{\bf #1} {(#2)} {#3}}}
\nc{\ijtp}[3]{{\it Int.\ J.\ Theor.\ Phys.\ }{{\bf #1} {(#2)} {#3}}}
\nc{\jmp}[3]{{\it  J.\ Math.\ Phys.\ }{{ \bf #1} {(#2)} {#3}}}
\nc{\jpa}[3]{{\it  J.\ Phys.\ A\ }{{\bf #1} {(#2)} {#3}}}
\nc{\jpc}[3]{{\it  J.\ Phys.\ C\ }{{\bf #1} {(#2)} {#3}}}
\nc{\jap}[3]{{\it J.\ Appl.\ Phys.\ }{{\bf #1} {(#2)} {#3}}}
\nc{\jpsj}[3]{{\it J.\ Phys.\ Soc.\ Japan\ }{{\bf #1} {(#2)} {#3}}}
\nc{\lmp}[3]{{\it Lett.\ Math.\ Phys.\ }{{\bf #1} {(#2)} {#3}}}
\nc{\mpl}[3]{{\it  Mod.\ Phys.\ Lett.\ }{{\bf #1} {(#2)} {#3}}}
\nc{\ncim}[3]{{\it  Nuov.\ Cim.\ }{{\bf #1} {(#2)} {#3}}}
\nc{\np}[3]{{\it  Nucl.\ Phys.\ }{{\bf #1} {(#2)} {#3}}}
\nc{\pr}[3]{{\it Phys.\ Rev.\ }{{\bf #1} {(#2)} {#3}}}
\nc{\pra}[3]{{\it  Phys.\ Rev.\ A\ }{{\bf #1} {(#2)} {#3}}}
\nc{\prb}[3]{{\it  Phys.\ Rev.\ B\ }{{{\bf #1} {(#2)} {#3}}}}
\nc{\prc}[3]{{\it  Phys.\ Rev.\ C\ }{{\bf #1} {(#2)} {#3}}}
\nc{\prd}[3]{{\it  Phys.\ Rev.\ D\ }{{\bf #1} {(#2)} {#3}}}
\nc{\prl}[3]{{\it Phys\ Rev.\ Lett.\ }{{\bf #1} {(#2)} {#3}}}
\nc{\pl}[3]{{\it  Phys.\ Lett.\ }{{\bf #1} {(#2)} {#3}}}
\nc{\prep}[3]{{\it Phys\. Rep.\ }{{\bf #1} {(#2)} {#3}}}
\nc{\prsl}[3]{{\it Proc.\ R.\ Soc.\ London\ }{{\bf #1} {(#2)} {#3}}}
\nc{\ptp}[3]{{\it  Prog.\ Theor.\ Phys.\ }{{\bf #1} {(#2)} {#3}}}
\nc{\ptps}[3]{{\it  Prog\ Theor.\ Phys.\ suppl.\ }{{\bf #1} {(#2)} {#3}}}
\nc{\physa}[3]{{\it  Physica\ A\ }{{\bf #1} {(#2)} {#3}}}
\nc{\physb}[3]{{\it  Physica\ B\ }{{\bf #1} {(#2)} {#3}}}
\nc{\phys}[3]{{\it Physica\ }{{\bf #1} {(#2)} {#3}}}
\nc{\rmp}[3]{{\it  Rev.\ Mod.\ Phys.\ }{{\bf #1} {(#2)} {#3}}}
\nc{\rpp}[3]{{\it Rep.\ Prog.\ Phys.\ }{{\bf #1} {(#2)} {#3}}}
\nc{\sjnp}[3]{{\it Sov.\ J.\ Nucl.\ Phys.\ }{{\bf #1} {(#2)} {#3}}}
\nc{\spjetp}[3]{{\it Sov.\ Phys.\ JETP\ }{{\bf #1} {(#2)} {#3}}}
\nc{\yf}[3]{{\it Yad.\ Fiz.\ }{{\bf #1} {(#2)} {#3}}}
\nc{\zetp}[3]{{\it Zh.\ Eksp.\ Teor.\ Fiz.\  }{{\bf #1}  {(#2)} {#3}}}
\nc{\zp}[3]{{\it Z.\ Phys.\ }{{\bf #1} {(#2)} {#3}}}
\nc{\ibid}[3]{{\sl ibid.\ }{{\bf #1} {#2} {#3}}}
\nc{\rf}[1]{(\ref{#1})}
\nc{\nn}{\nonumber \\*}
\nc{\bfB}{\bf{B}}
\nc{\bfv}{\bf{v}}
\nc{\bfx}{\bf{x}}
\nc{\bfy}{\bf{y}}
\nc{\vx}{\vec{x}}
\nc{\vy}{\vec{y}}
\nc{\oB}{\overline{B}}
\nc{\oI}{\overline{I}}
\nc{\oR}{\overline{R}}
\nc{\rar}{\rightarrow}
\nc{\ti}{\times}
\nc{\slsh}{\hskip-5pt/}
\nc{\sm}{Standard~Model~}
\nc{\MP}{M_{\rm Pl}}
\nc{\tp}{t_{\rm Pl}}
\nc{\ave}{\bar{E}}

\renc{\min}{p_{\rm min}}
\renc{\max}{p_{\rm max}}
\nc{\pmin}{p_{\rm min}}
\nc{\pmax}{p_{\rm max}}
\nc{\fo}{f_0}
\nc{\foi}{f_{0,i}\,}
\nc{\fop}{f_0^P}
\nc{\fou}{f_0^U}
\def\sepand{\rule{14cm}{0pt}\and}
\nc{\eff}{{\rm eff}}
\nc{\MT}{M_{\rm T}}
\nc{\ML}{M_{\rm L}}
\nc{\kk}{\vek{k}}
\nc{\pp}{{\rm p}}
\nc{\cb}{critical bubble~}
\nc{\cbs}{critical bubbles~}
\nc{\scb}{subcritical bubble~}
\nc{\scbs}{subcritical bubbles~}
\nc{\mm}{m_0}
\nc{\gluino}{m_{\rm gluino}}
\begin{document}

{\title{{\vspace{-1.2cm}\hfill {{\small HU-TFT-95-38\\
         \hfill HU-SEFT R 1995-11\\
        }}\vskip 1truecm}
{\bf R-parity breaking phase transition in the susy singlet majoron model}}

\vspace{-1.2cm}

\author{
{\sc Kari Enqvist$^{1}$} \\
{\sl\small Department of Physics, P.O. Box 9, FIN-00014
University of Helsinki, Finland}\\
{\sc Katri Huitu$^{2}$}\\
{\sl\small Research Institute for High Energy Physics}\\
{\sl\small P.O. Box 9, FIN-00014 University of Helsinki, Finland}\\
and\\
{\sc P.N. Pandita }\\
{\sl\small Department of Physics,}
{\sl\small North Eastern Hill University, P.O. Box 51, Laitumkhrah} \\
{\sl\small Shillong 793 003, India} \\
\sepand
}
\maketitle}
\vspace{1.0cm}
\begin{abstract}
\noindent
We consider thermal properties of the susy singlet majoron model.
We compute the critical temperature $T_c$ and the subsequent reheating
temperature $T_{\rm RH}$ for $R$-parity breaking. Succesful baryogenesis
constrains the parameter space via the requirements that $T_c$ and $T_{\rm RH}$
are lower than the electroweak phase transition temperature. A further
constraint is provided by requiring that the gauge singlet should be
kinematically allowed to decay, in order not to have a matter dominated
universe at the time of nucleosynthesis. We have made a detailed study
of the parameter space and find an upper limit for the susy breaking
scalar mass $\mm\lsim 750~(900)$ GeV if $\gluino=100~(1000)$ GeV,
which is valid except for certain special values of the singlet sector
parameters.

\end{abstract}
\vfill
\footnoterule
{\small  $^1$enqvist@pcu.helsinki.fi; $^2$huitu@phcu.helsinki.fi}
\thispagestyle{empty}
\newpage
\setcounter{page}{1}
In the minimal supersymmetric standard model (MSSM), gauge invariance and
supersymmetry allow terms in the superpotential which violate baryon($B$) or
lepton($L$) number. The absence of these dangerous terms is ensured by assuming
a conserved discrete symmetry, called $R$-parity, $R_p=(-1)^{3B+L+2S}$, where
$S$ is the spin of the particle \cite{rp}. Because of its connection to lepton
number, it is interesting to consider situations where $R$-parity is slightly
broken, either explicitly \cite{excplitly} or spontaneously \cite{spont}.
Spontaneuous breaking of lepton number is possible in MSSM without introducing
additional fields, since the scalar partner of the neutrino may acquire a
non-vanishing vacuum expectation value \cite{spont}, giving rise to a majoron
\cite{maj1,maj2}, which in this case is mainly the supersymmetric
 partner of the neutrino. Since such a majoron should be detected in $Z^0$
decay, the measurements of the $Z^0$ width \cite{width} rule out this
possibility, unless one assumes that there is small explicit {\it L} violation,
in addition to the spontaneous one, in the MSSM \cite{small}. On the other hand
it is possible to break $R$-parity  spontaneously in the MSSM with the
inclusion of additional singlet superfields \cite{singlets}, such that the
resulting majoron, which is now dominantly a singlet under the gauge
group, is not in conflict with the measurements of the $Z^0$ width.
A particularly attractive scheme which implements such a spontaneous
violation of  $R$-parity is the supersymmetric version \cite{susy1,susy2}
of the singlet majoron model. This model is the simplest extension which
incorporates successfully the  spontaneous $L$-(and $R$-parity-)violation
in the MSSM. In this model there are, besides the chiral superfields
of the MSSM,  right-handed neutrino chiral superfields $N_i$ (one for
each generation) and an additional gauge singlet
superfield $\Phi$, having two units of $L$. The superpotential
of the susy singlet majoron model, which is invariant under
gauge symmetry and $L$ is, in the standard notation,
\be{spot}
W=h^UQU^cH_2+h^DQD^cH_1+h^ELE^cH_1+h^\nu LNH_2+\mu H_1H_2+\lambda NN\Phi~,
\ee
where we have suppressed the generation indices. \eq{spot} contains the
usual terms of MSSM together with the Yukawa interactions for the
right-handed neutrinos $N_i$ and an interaction term for the gauge singlet
$\Phi$. It has been shown, through an analysis of the renormalization
group equations (RGE), that for a wide range of parameters one
can obtain radiative breaking of  $R$-parity in this model \cite{susy1,susy2}.
It has also been argued that the  $L$-(and $R$-parity-)-violating
transition may take place after the electroweak phase transition.
In that case the sphaleron-induced $B$-violating transitions are frozen
out, so that any pre-existing baryon asymmetry is unaffected by this
phase transition.

In this paper we extend and improve the analysis of ref. \cite{susy1} and
 study the $L$-breaking cosmological phase transition of the
susy singlet majoron model in detail.
In particular, we compute the
critical temperature $T_c$ and the reheating temperature $T_{\rm RH}$
and compare them with the critical temperature of the electroweak phase
transition. We also point out that the gauge singlet sector should decay
fast enough in order not to clash with primordial nucleosynthesis. All the
constraints, when taken together, imply that succesful baryogenesis
in the susy singlet Majoron model requires an upper limit on the
susy breaking scalar mass $\mm$.

To start, we recall the tree level potential for the susy singlet majoron model
\cite{susy1,susy2}. It consists of three pieces, and can be written along
the neutral directions as
\be{potential}
V=V_H(H^0_1,H^0_2)+V_{{\tilde N}\Phi}({\tilde N}_i,\Phi)+
V_{\tilde\nu}(\tilde\nu_i,{\tilde N}_i,\Phi,H_1^0,H^0_2)~.
\ee
Here, $V_H$ is the Higgs potential of the MSSM. We assume that
the couplings $\lambda_{ij}$ are real and work in the basis
where they are diagonal, in which case we can write
the potential ($\lambda_{ij} = \lambda_i \delta_{ij}$)
\be{nphi}
V_{{\tilde N}\Phi}=\sum_i m_{{\tilde N}_i}^2\vert {\tilde N}_i\vert^2
+m_\Phi^2\vert\Phi\vert^2-(\sum_iA_i\lambda_i{\tilde N}_i^2\Phi+h.c.)
+4\sum_i\vert \lambda_i{\tilde N}_i\Phi\vert^2+\vert\sum_i\lambda_i
{\tilde N}_i^2\vert^2~,
\ee
where $m_{{\tilde N}_i}^2$ and $m_\Phi^2$ are soft susy breaking masses,
and $A_i$ are
the trilinear couplings. Finally, if we assume that the coupling constants
$h^\nu$ are small, and retain only the leading terms in $h^\nu$, then we
may write the third piece of the potential \eq{potential} as
\bea{nupot}
V_{\tilde\nu}&=&\sum_i m_{\tilde\nu_i}^2\vert \tilde\nu_i\vert^2
+\left[\sum_{ij}h^\nu_{ij}\tilde\nu_i(
2\lambda_j{\tilde N}_j^*\Phi^*H_2^{0}-\mu {\tilde N}_jH_1^{0*}-
A^{(h)}_{ij}{\tilde N}_jH^0_2)+h.c
\right] \\
&+&\frac 18(g^2+{g´}^2)\left[(\sum_i\vert\tilde\nu_i\vert^2)^2
+2\sum_i\vert\tilde\nu_i\vert^2
(\vert H_1^0\vert^2-\vert H_2^0\vert^2)\right]~,
\eea
where $m_{\tilde\nu_i}^2$ are the soft susy breaking masses, $A^{(h)}_{ij}$
are the trilinear couplings and $g$ and $g'$ are the SU(2) and U(1)
gauge couplings, respectively. Assuming, as in the
case of MSSM, that all susy breaking masses and trilinear
couplings are equal to a universal mass $\mm\sim 10^2-10^3$ GeV and a
universal coupling $A$, respectively, at some GUT scale $M_U\sim 10^{16}$ GeV,
the values of the parameters in the scalar potential can be obtained by
solving the appropriate RGEs. The effect of running of these parameters
is to drive SU(2)$\times$U(1) breaking through $V_H(H_1^0,H_2^0)$ with
$\langle H^0_1\rangle\equiv v_1=v\cos\beta$ and
$\langle H_2^0\rangle\equiv v_2=v\sin\beta$, where
$v=v_1^2+v_2^2=(174~\GeV)^2$. Furthermore, for a wide range of parameters
the nontrivial global minimum of $V_{{\tilde N}\Phi}({\tilde N}_i,\Phi)$
is realized in
such a manner as to break the global lepton number and $R$-parity:
\be{vevs}
\langle\Phi\rangle=\phi,~~\langle {\tilde N}_i\rangle=y_i ¨,
\ee
where typically $\phi\sim y_i\sim\mm$. Then, nonzero vevs are induced
for the sneutrinos
through the $h^\nu$-coupling which connects the ordinary doublet Higgs and
lepton sectors to the singlet sector.
If $m_{{\tilde\nu}_i}\sim\mm,~\phi\sim y_i\sim\mu\sim\mm,~v_1\sim v_2\sim M_W$
 and
$A^{(h)}\sim \mm$, then $\langle{\tilde\nu}_i\rangle\sim h^\nu  M_W$.

The vevs $\phi$ and $y_i$ are determined from the minimum of the potential
$V_{{\tilde N}\Phi}$. We assume that
$\lambda_3\gg \lambda_1,~\lambda_2$ (a minimum
with $\lambda_3= \lambda_1=\lambda_2$ is not favoured by the solution
of RGEs \cite{susy1,susy2}). The minimum at scale $Q$ is defined by
\bea{minimum}
\phi&=&{x\over 4\lambda_3}~, ~~y_3^2={m_\Phi^2\over 4\lambda_3^2}
{x\over (A_3-x)}~,~~y_2=y_1=0~,\\
0&=&x^3-3A_3x^2+2(m_{{\tilde N}_3}^2-m_\Phi^2+
A_3^2)x-4A_3m_{{\tilde N}_3}^2~,~~A_3/x>1~.
\eea
Lepton number is broken spontaneously if \eq{minimum} yields the absolute
minimum, which is obtained when
\be{absol}
A_3-m_\Phi<x<A_3+m_\Phi~,~~m^2_\Phi>0~.
\ee
Assuming universal boundary conditions at the GUT scale $M_U$, the
parameters of the potential $V_{N\Phi}$ at the low scale $Q$ can be
written as \cite{susy1,susy2}
\bea{masses}
m_\Phi^2(Q)=\frac 15\mm^2[2+(3-A^2)K^2+A^2K^4]~,\\
m^2_{{\tilde N}_{1,2}}=\mm^2~,~~m^2_{{\tilde N}_3}(Q)=2m_\Phi^2(Q)-\mm^2~,\\
A_{1,2}(Q)=AK^{\frac 25}~,~~A_3(Q)=AK^2~,\\
\lambda_{1,2}(Q)=0~,~~\lambda_3(Q)=\lambda_0K~,
\eea
where
\be{running}
K=\left[1+{5\over 4\pi^2}\lambda_0^2\ln\left({M_U\over Q}\right)\right]
^{-\frac 12}~.
\ee
{}From the scalar potential, \eq{potential}, one can derive the
mass squared matrix for
the scalar bosons \cite{pandita} in a straightforward manner.
In this paper we shall only
consider the situation when $\lambda_3$ is the largest of the non-zero
values of $\lambda_i$, so that \eq{minimum} holds. In the limit
$h^\nu\to 0$ the mass squared matrix is of a block diagonal form
${\cal M}_S^2=diag(M_H^2,M^2_{\tilde{\nu}}, M^2_{{\tilde N}\Phi})$,
where $M_H^2$
is effectively the $2\times 2$ mass squared matrix of the scalar Higgs
bosons in MSSM \cite{gunion}, $M^2_{\tilde{\nu}}$ is the $3\times 3$
mass squared matrix of left-handed sneutrinos, and
\be{nfii}
M^2_{{\tilde N}_3\Phi}=\pmatrix{{A_3\lambda_3y_3^2
\over\phi}& 8\lambda^2_3y_3\phi-
2A_3\lambda_3y_3\cr
  8\lambda^2_3y_3\phi-2A_3\lambda_3y_3 & 4\lambda^2_3y^2_3}~.
\ee

 To study the $L$-breaking phase transition, we shall focus on the
$({\tilde N}_3,\Phi)$-sector only. At high temperatures ${\tilde N}_3$ and
$\Phi$ are brought
into equilibrium by decays, inverse decays and ordinary scattering
processes.
The thermally averaged rate for a given process is, neglecting
final state blocking,   given by
\bea{gamma}
\Gamma&=&\int\prod_k d\Pi_k(2\pi)^4\delta(P_i-P_f)\prod_{\rm initial}
{f_i\over n_i}\vert M\vert^2\\
&=&{1.4\times 10^{-2}T^{-1}\vert M\vert^2}~~~~~(1\to 2+3)~,\\
&=&4\times 10^{-4}T\vert M\vert^2~~~~~(2\to 2)~,
\eea
where $d\Pi_k=d^3p_k/(2\pi)^32E_k$, $f_i$ is the momentum distribution
of the particles in the initial state, and $n_i$ is their number density.
In Eqs. (17) and (18)  a constant matrix element is assumed.
These rates are to be compared with the Hubble rate $H\simeq 23T^2/M_{Pl}$.
{}From \eq{gamma}
one easily finds that for $\lambda$ not too small, ${\tilde N}_3$
and $\Phi$ are in
equilibrium already at temperatures much above 1 TeV, because of the
reactions ${\tilde N}_3\Phi\to {\tilde N}_3\Phi$ and $\Phi\to {\tilde N}_3
{\tilde N}_3$. The fields of the MSSM
can also be assumed to be in equilibrium already at high temperatures.

On the other hand, it is not obvious that the $({\tilde N}_3,\Phi)$-system
is in
thermal contact with $H_1^0,~H_2^0$ and $\tilde\nu$. This is because the
coupling is only through $h^\nu$, which is small. Here the most
important reaction is the decay $H_1^0\to\tilde\nu_i {\tilde N}_j$, which
induces chemical equilibrium at temperatures
\be{limit}
T\lsim 194\vert{\mu\over\GeV} h^\nu_{ij}\vert^{2/3}~\TeV~.
\ee
If $\mu\simeq 100$ GeV, there is no thermal contact at the
electroweak phase transition if $h^\nu\lsim 10^{-7}$. This would mean that
the $(N_3,\Phi)$-system and the MSSM would experience different temperatures.
In what follows we tacitly assume that this is not the case.
For light $\Phi,~{\tilde N}_3$ and $H_2^{\pm}$ also the scattering process
$\Phi H_2^+\to {\tilde N}_3e_L$ is important.
For instance, at $m/T\simeq 0.1$ the
particles are kept in equilibrium at temperatures
$T\lsim 10^{22}\lambda_3^2{h^{\nu}}^2$. However, for increasing particle
masses the thermalization temperature decreases rapidly.

At high temperatures the plasma masses
of ${\tilde N}_3$ and $\Phi$ are given by \cite{susy1}
\be{thermal}
M^2_\Phi(T,Q)=m_\Phi^2(Q)+\frac 12\lambda^2 (Q)T^2~;~~
M^2_{\tilde N}(T,Q)=m_{\tilde N}^2(Q)+\lambda^2 (Q)T^2~,
\ee
where the RGE masses $m^2(Q)$ are given by \eq{masses}, and we
have neglected terms of order $\cO (h^\nu)$. Here $Q$ is the
renormalization point, which in principle is independent of $T$. However,
as in thermal bath the average momenta of the particles are peaked about
$3T$, it makes sense to choose the renormalization point to depend on
temperature, so that $Q\simeq T$. Setting $Q\simeq m_{\rm susy}\simeq m_0$ as
usual would
yield masses which would differ by terms of the order of
${\cal O}(\lambda)$; for our purposes
this difference is inessential. With this choice, we may use
\eq{thermal} to search
for the critical temperature $T_c$, at which the high $T$ minimum
$({\tilde N}_3,\Phi)=(0,0)$ becomes unstable.

Note that the expressions \eq{thermal} are valid
if the temperature is much bigger
than the masses circulating in the loops. There is a subtlety here in
that at some point $m_{\tilde N}^2(Q)$ becomes negative, and the finite $T$
perturbation expansion, leading to \eq{thermal},  becomes unstable.
This can be remedied by performing resummation in the graph involving
${\tilde N}_3$ loop, which effectively replaces  $m_{\tilde N}^2(Q)$ by
the plasma mass
 $M_{\tilde N}^2$, but does not change the potential to lowest
order in $\lambda$.
As  $M_{\tilde N}^2\to 0$ for $T\to T_c$, the plasma mass of ${\tilde N}_3$
may always be
assumed to be less than $T$. Hence loops of ${\tilde N}_3$ always contribute to
\eq{thermal}.

The resummed loop involving  $\Phi$ becomes Boltzmann-suppressed when
$M_\phi\gsim T$. This implies that if $\mm^2\gsim 2T^2_c$, loops of
$\Phi$ should be neglected. Effectively, this means that
$\lambda^2 T^2\to \frac 12 \lambda^2 T^2$ in  $M_{\tilde N}^2$ in \eq{thermal}.
The critical temperature for {\it L}-breaking is, however, in both
cases defined through  $M^2_{\tilde N}(T_c,T_c)=0$.

For a given $A$ and $\lambda$, it is then possible to run the
RGEs to find $T_c$ using \eq{thermal}. In addition, one has to
impose the conditions of \eq{minimum} to make sure that a zero temperature
global minimum exists for the chosen set of parameters. We have done
this numerically, and the result, displaying the allowed region for
fixed $T_c/\mm$, is shown in Fig. 1.

The parameter space can be  constrained by noting that the physical
eigenstates of the $({\tilde N}_3,\Phi)$-sector
should be unstable. Otherwise the energy density $\rho_{osc}$ associated with
coherent field oscillations about the vacuum  would soon
after the $R$-breaking phase transition start to
dominate the energy density of the universe, with disastrous consequences
for e.g. primordial nucleosynthesis. Roughly, $\rho_{osc}\simeq m^2v^2
(T/T_c)^3$ where $m$ and $v$ are a generic mass and vev, respectively.
Thus, in the absence of dissipation,  $\rho_{osc}$ would soon become
larger than radiation energy density $\rho_{rad}\sim T^4$.

The eigenstates can easily be found
from \eq{nfii} for a given  value of $\lambda$ and $A$. It turns out that
the heavier eigenstate, with mass
$m_2$, is predominantly $\Phi$. Because $\Phi$ does not couple directly to
the MSSM states,
it decays either via the small mixing with ${\tilde N}_3$
or via an ${\tilde N}_3$ virtual state. Here we shall assume that the latter is
the case.
Requiring that the process $\Phi\to {\tilde N}_3
{\tilde N}_3\to
{\tilde N}_3\nu_L{\tilde\chi}_i^0$, where $\chi_i^0$ is the lightest
neutralino, is not kinematically forbidden implies
$m_{\Phi}-m_{{\tilde N}_3}>m_{{\tilde\chi}_i^0}>$ 18 GeV, where the last figure
is the experimental lower bound on LSP \cite{width}. The mass difference
$m_{\Phi}-m_{\tilde N}$ is a function of $m_0, A$ and $\lambda$, and
the different contours are  displayed in Fig. 1. If the gluino is
light, $\gluino\simeq 100$ GeV, the experimental lower limit on
sneutrino, $m_{\tilde \nu}>37.1$ GeV \cite{width}, rules out part of
the parameter space as indicated in Fig. 1.

Note that the actual decay rate depends on the $\Phi-{\tilde N}_3$
mixing and the
mixing of the gauge singlets with the MSSM sparticles, which is
characterized by the unknown (but small) coupling $h^\nu$. We have
checked that  $\Phi-{\tilde N}_3$ mixing is indeed small in the physical
large $\lambda$ region. It is conceivable, though, that for some
values of $h^\nu$ the heaviest eigestate could decay directly to
MSSM particles already before the onset of nucleosynthesis. In any case
there always is some kinematic constraint that must be
satisfied in order that the decay is possible.

The critical temperature for $L$-breaking can further be constrained by
baryogenesis considerations. $L$-violating interactions, if in equilibrium
with the anomalous elecroweak $B+L$-violating interactions,
will wash out any  pre-existing baryon asymmetry. This can be avoided if
$R$-parity breaking couplings are very small, so that these interactions
are never in equilibrium. In this case $R$-parity breaking would not be
relevant for experiments. Another possibility, as suggested in
\cite{susy1}, is to impose the condition that $L$-breaking
takes place after electroweak phase transition, so that the sphaleron
induced transitions have already dropped out of equilibrium.

 In the
small $h^\nu$ approximation adopted in this paper, the critical temperature
for the electroweak phase transition may be assumed to be the same as
in the MSSM. Computing
the critical temperature is, however, notoriously difficult task, and
it is conceivable that it is dominated by non-perturbative effects like
in the non-supersymmetric case \cite{nonpert}.
Because the transition is presumably
only weakly first order, and the latent heat release is small, for our
purposes it suffices to estimate the electroweak critical temperature
by the spinoidal instability temperature $T_0$, determined by
$det(M^2_H(T_0))=0$. The resummed one-loop  mass matrix $M^2_H$ has been
presented in \cite{mmatrix}. The expressions are lenghty, and we do
not reproduce them here. Using these, we have varied the parameters of the MSSM
in the following ranges: $1<\tan\beta<50$,
the trilinear coupling associated with the stop-sector
$0<A_t<1~\TeV$ and the  Higgs mixing $0<\mu<1~\TeV$.
We have explicitly studied two cases, the case of a light gluino with
$\gluino=100$ GeV, and the case of a heavy gluino, $\gluino=1$ TeV.
The resulting $T_0$ is shown in Fig. 2.

Succesful baryogenesis requires that $T_c<T_0$. Moreover, we should also
require that the reheating temperature $T_{\rm RH}$ after $R$-parity
breaking does not exceed $T_0$ so as to make sphaleron interactions
operative again \cite{susy1}. The reheating temperature is given by
\be{reheat}
T_{\rm RH}\simeq \left({30\vert V_{\rm min}\vert\over \pi^2g_*}\right)^{\frac
14}
\ee
where $V_{\rm min}$ is the global minimum and we have taken
$g_*\simeq 100$. In Fig. 2 we show $T_{\rm RH}$
for a choice of parameter values. In Fig. 3 we have taken all the constraints
together to find the allowed parameter space.
If $A\gsim 2$, we find the upper limits
\be{raja}
\mm\lsim 750~\GeV~~(\gluino=100~\GeV);~~\mm\lsim 900~\GeV~~(\gluino=1~\TeV)~,
\ee
where the figures refer to the case $\tan\beta=1.5$; if $\tan\beta=50$,
the limits are slightly relaxed, as can be seen in Fig. 3.
However, the limits \eq{raja} are not completely general. If it happens
that $A\lsim 2$ and $\lambda\simeq 1$, $m_0$ can be as high as 4 TeV. Note
that for $\lambda\ll 1$ one  gets $A\simeq 3$ so that \eq{raja}
always holds. We wish to emphasize that in all the cases an upper limit on
$m_0$ exists.

To conclude, we have studied  in detail the cosmological
$R$-parity breaking phase
transition in the susy singlet majoron model. In particular,
we focussed on the constraints imposed on the model by the requirements
of a succesful baryogenesis. The necessary conditions are that
$T_c$ and $T_{\rm RH}$ are lower than the electroweak phase transition
temperature $T_0$, which depends in a complicated way on the parameters
of MSSM. The connection to the gauge singlet sector of the majoron model
is provided by the common scalar mass parameter $\mm$. An additional
cosmological constraint is that the gauge singlet sector should be
allowed to decay. This is a kinematical constraint that restricts the
range of $\mm$. Because susy sparticle spectrum is essentially given
by $\mm$, plus D-term and radiatice corrections,
 baryogenesis requires a sparticle
spectrum in the susy singlet majoron model which should be observable
at LHC.

\vspace{1.cm}
\noindent
{\bf Acknowledgements}

\vspace{1.cm}
\noindent
One of the authors (PNP) is grateful to NORDITA and Research Institute
of Theoretical Physics, University of Helsinki, for the hospitality
during the course of this work.
The work of PNP was supported by the Department of Atomic Energy,
India.
\newpage
%

\input{epsf.sty}
\begin{figure}[t]
\leavevmode
\begin{center}
\mbox{\epsfxsize=15.cm\epsfysize=15.cm\epsffile{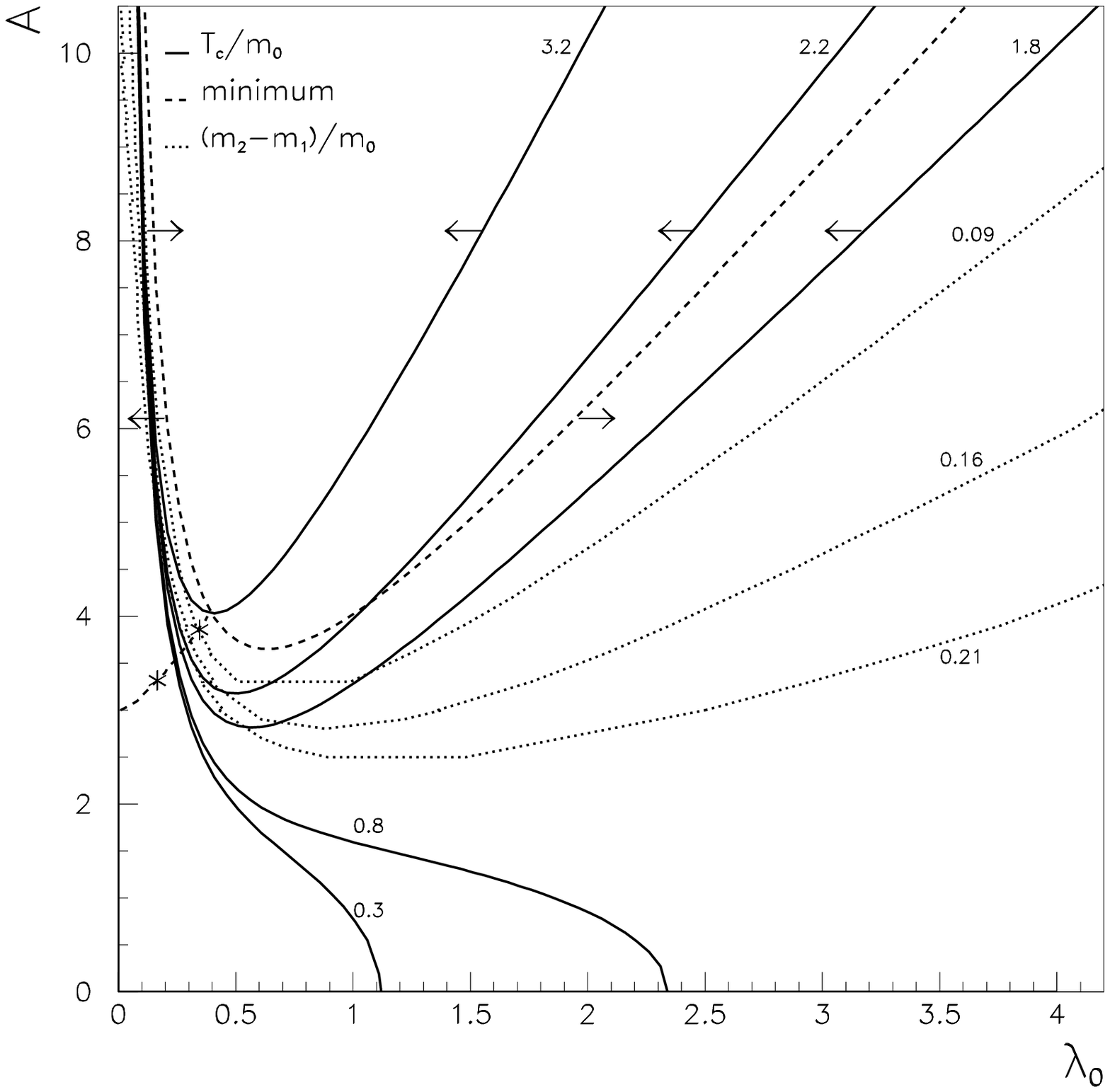}}
\end{center}
\caption{
The allowed values for $A$ and $\lambda_0$.
The solid lines indicate the parameter space for which R-parity is
not broken at temperatures $T>T_c$ with
T$_{c}$/m$_0$ values shown in the figure.
The dashed line indicates the parameter space allowed by the
requirement of global minimum.
The dotted lines are contours of $m_2-m_1/m_0$,
where $m_{1,2}$ are the physical eigenstates of $\Phi$ and
$\tilde N_3$.
{}From experimental lower limit for $\tilde\nu$, the area above the
starred-dashed line is not allowed if
$m_{gluino}\sim $ 100 GeV, $\mu =A_t=0$ and tan $\beta$=1.5.}
\label{fig1}
\end{figure}

\begin{figure}[t]
\leavevmode
\begin{center}
\mbox{\epsfxsize=15.cm\epsfysize=15.cm\epsffile{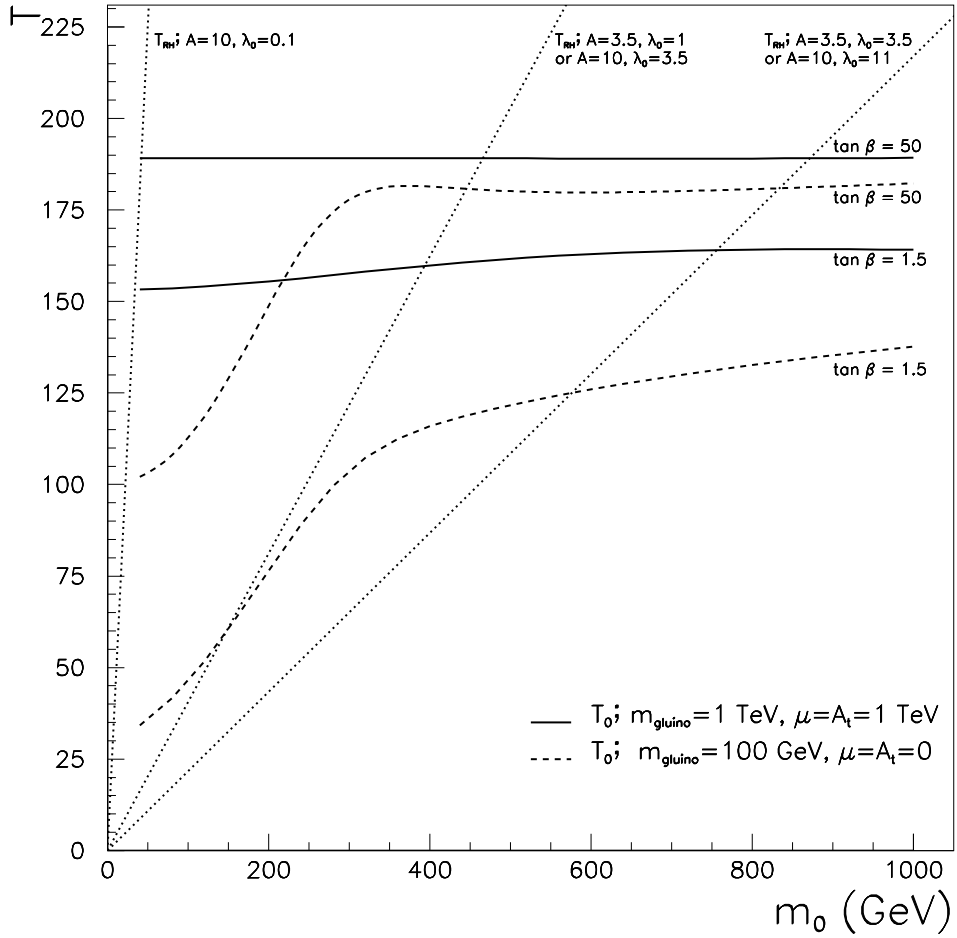}}
\end{center}
\caption{
Spinoidal instability temperature $T_0$ (solid and dashed lines)
and reheating temperature $T_{RH}$ (dotted lines)
as functions of $m_0$.
The solid lines correspond to $m_{gluino}=1$  TeV,
$\mu =A_t =1$ TeV,
and the dashed ones to $m_{gluino}=$ 100 GeV,
$\mu =A_t =0$.
As indicated, $\tan\beta=1.5$ or 50.
}
\label{fig2}
\end{figure}

\begin{figure}[t]
\leavevmode
\begin{center}
\mbox{\epsfxsize=15.cm\epsfysize=15.cm\epsffile{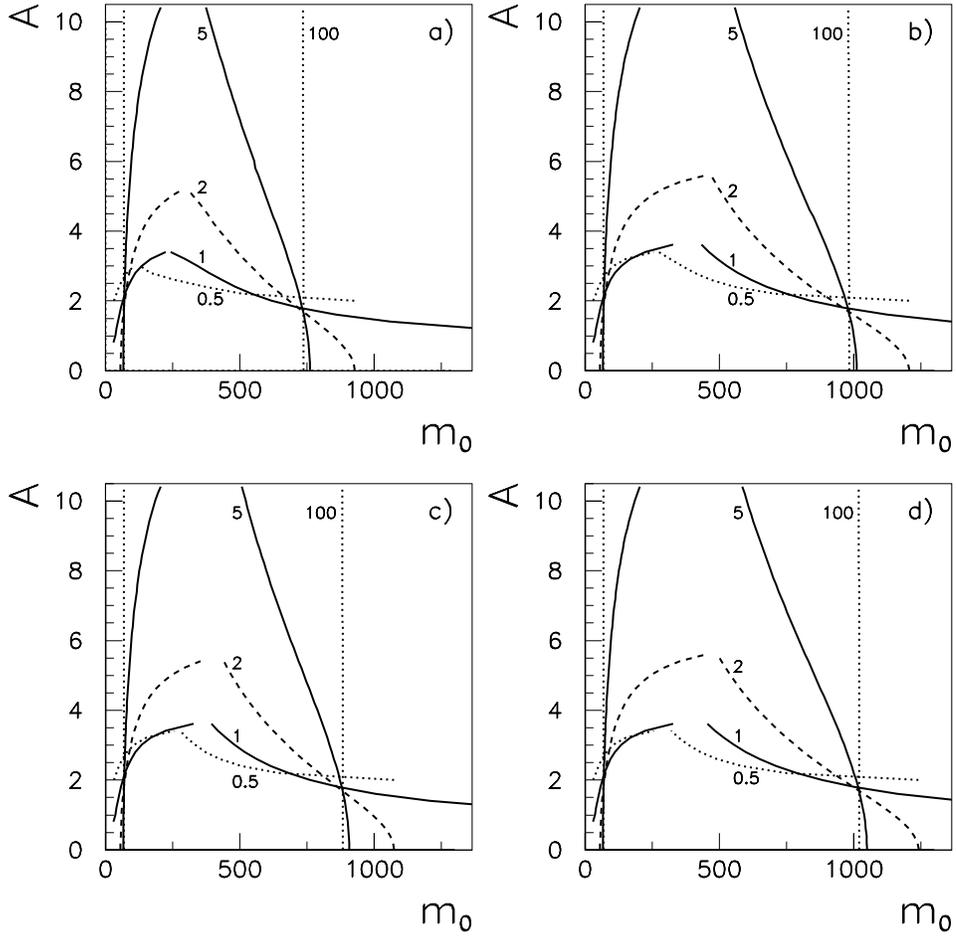}}
\end{center}
\caption{
The allowed values of $m_0$ and $A$
for $\lambda_0$ values 0.5, 1, 2, 5, and 100.
The allowed range for $m_0$,  which is the area between the lines,
is shown for the cases
$m_{gluino}=$ 100 GeV, $\mu =A_t =0$ with
a) tan $\beta=1.5$,
b) tan $\beta=50$, and for
$m_{gluino}=$ 1 TeV, $\mu =A_t =1$ TeV with
c) tan $\beta=1.5$ and
d) tan $\beta=50$.
}
\label{fig3}
\end{figure}

\end{document}